\documentclass[12pt]{article}
\usepackage{amssymb}
\usepackage{amsmath}
\def\bz{\bar{z}}

\def\be{\begin{equation}}
\def\ee{\end{equation}}
\def\arr{\begin{array}{rll}}
\def\ea{\end{array}}
\def\bea{\begin{eqnarray}}
\def\eea{\end{eqnarray}}

\def\R{{\mathbb R}}

\def\ic{{\rm i}}
\def\eu{{\rm e}}
\def\N2{$N{=}2$}
\def\pa{\partial}
\def\diff{{\rm d}}

\def\sfrac#1#2{{\textstyle\frac{#1}{#2}}}
\def\>{\rangle}
\def\<{\langle}
\def\+{\dagger}
\def\={\ =\ }
\def\and{\quad\textrm{and}\quad}
\def\und{\qquad\textrm{and}\qquad}
\begin{document}
\renewcommand{\thefootnote}{\fnsymbol{footnote}}

\begin{titlepage}
\setcounter{page}{0}
\begin{flushright}
LMP-TPU--7/13  \\
ITP-UH--12/13
\end{flushright}
\vskip 1cm
\begin{center}
{\LARGE\bf On two-dimensional integrable models}\\
\vskip 0.5cm
{\LARGE\bf with a cubic or quartic integral of motion}\\
\bigskip
\vskip 1cm
$
\textrm{\LARGE Anton Galajinsky\ }^{a} \quad \textrm{\Large and} \quad
\textrm{\LARGE Olaf Lechtenfeld\ }^{b}
$
\vskip 1cm
${}^{a}$ {\it
Laboratory of Mathematical Physics, Tomsk Polytechnic University, \\
634050 Tomsk, Lenin Ave. 30, Russian Federation} \\[4pt]
{Email: galajin@tpu.ru}
\vskip 0.5cm
${}^{b}$ {\it
Institut f\"ur Theoretische Physik und Riemann Center for Geometry and Physics, \\
Leibniz Universit\"at Hannover, Appelstrasse 2, 30167 Hannover, Germany} \\[4pt]
{Email: lechtenf@itp.uni-hannover.de}
\end{center}
\vskip 2cm
\begin{abstract}
\noindent
Integrable two-dimensional models which possess an integral of motion cubic or quartic
in velocities are governed by a single prepotential, which obeys a nonlinear partial
differential equation. Taking into account the latter's invariance under continuous
rescalings and a dihedral symmetry, we construct new integrable
models with a cubic or quartic integral, each of which involves either one or two
continuous parameters. A reducible case related to the two-dimensional wave equation
is discussed as well. We conjecture a hidden $D_{2n}$ dihedral symmetry for models
with an integral of $n$th order in the velocities.
\end{abstract}
\vfill
\end{titlepage}

\renewcommand{\thefootnote}{\arabic{footnote}}
\setcounter{footnote}0

\noindent
{\bf 1. Introduction and setting of the problem}

\vskip 0.3cm
\noindent
The study of two-dimensional integrable models possessing an integral of motion
which is cubic or quartic in velocities has a long history.
Having been discovered originally within the context of rigid body dynamics~\cite{SK,Gor},
such systems were later generalized in a number of ways and have been a topic
of active research for the last three decades \cite{Henon}--\cite{YH}
(for a review see~\cite{H}).

A conventional two-dimensional conservative system is governed by the Lagrangian
\be\label{cs}
L\=\frac 12 \sum_{i,j=1}^2 a_{ij}(q)\,{\dot q}^i {\dot q}^j-V(q)\ ,
\ee
where $a_{ij}(q)$ is assumed to be an invertible matrix which encodes the geometry
of a curved background, and $V(q)$ is a potential.
Because $a_{ij}(q)$ and $V(q)$ do not explicitly depend on time, the energy is conserved.
Given the potential $V(q)$ and the metric $a_{ij}(q)$,
several methods to uncover a second integral of motion have been proposed.
In particular, these include the Painlev\'e analysis, the Lax-pair approach and
the separation of variables in the Hamilton--Jacobi equation.
A more direct method considers a polynomial $I_2$ in the velocities,
with the coefficients being arbitrary functions of the coordinates,
and requires that it be conserved in time.
This condition yields a system of coupled nonlinear partial differential equations
on the coefficients, which also involve $a_{ij}(q)$ and $V(q)$.

In general, by applying a coordinate transformation and a redefinition of
the temporal coordinate, one can reduce the number of coefficients in the polynomial
and, thus, simplify the system of partial differential equations.
As was demonstrated in~\cite{Y6}, 
one can always choose new coordinates, $q^i=q^i(x,y)$ such that the system in terms of $(x,y)$
is on the zero energy level,\footnote{
Here and in what follows we use the conventions in \cite{Y6,Y}. 
The subscripts denote partial derivatives with respect to the corresponding variable. 
Throughout the paper we impose time-reversal invariance.}
\be\label{EqM}
\ddot x\=U_x \qquad
\ddot y\=U_y\ , \qquad I_1\ \equiv\ {\dot x}^2+{\dot y}^2-2 U\=0\ .
\ee

For the case of a conserved polynomial cubic in the velocities, 
the new variables simplify this integral to
\be\label{I2cub}
I_2\={\dot x}^3 + J(x,y)\,\dot{x} + K(x,y)\,\dot{y}\ ,
\ee
involving only two functions $J(x,y)$ and $K(x,y)$ to be determined.
Taking into account the equations of motion and the zero energy condition (\ref{EqM}),\footnote{
When computing the derivative of $I_2$, a term which involves ${\dot y}^2$ appears. 
In that term ${\dot y}^2$ should be changed by ${\dot y}^2=2 U-{\dot x}^2$, 
which then leads to (\ref{KJU}).} 
one can verify that $I_2$ is conserved in time provided the system of partial differential equations
\be\label{KJU}
K_x+J_y\=0\ , \qquad J_x-K_y+3 U_x\=0\ ,\qquad U_x J+U_y K+2 K_y U\=0\
\ee
holds \cite{Y6}.

Without loss of generality one can choose $K(x,y)$ in the form
\be\label{K}
K(x,y)\=-E_{xy}(x,y)\ ,
\ee
where $E(x,y)$ is a function to be determined. Then the leftmost equation in (\ref{KJU}) implies
\be\label{J}
J(x,y)\=E_{xx}(x,y)+\alpha(x)\ ,
\ee
where $\alpha(x)$ is an arbitrary function of $x$ only. As is evident from (\ref{K}) and (\ref{J}), 
$\alpha(x)$ can always be removed by a redefinition of $E(x,y)$. So in what follows we will ignore it. 
Substitution of (\ref{K}) and (\ref{J}) into the second equation in (\ref{KJU}) yields
\be
E_{xx}+ E_{yy}+3 U=\beta(y)\ ,
\ee
where $\beta(y)$ is an arbitrary function of $y$ only. 
Again, one can get rid of $\beta(y)$ by a redefinition of $E(x,y)$, 
which does not alter the form of $K(x,y)$ or $J(x,y)$ while fixes the potential
\be
U\=-\sfrac 13 \bigl( E_{xx}+ E_{yy}\bigr)\ .
\ee
The rightmost equation in (\ref{KJU}) then reads~\cite{Y6} (see also a related earlier work~\cite{LH})
\be\label{ME}
E_{xx} (E_{xxx}+E_{xyy})-E_{xy} (E_{yxx}+E_{yyy})-2 E_{xyy} (E_{xx}+E_{yy})\=0\ .
\ee

To summarize, the dynamical system (\ref{EqM}), 
which admits an integral of motion cubic in the velocities, 
is governed by the triple
\be
J(x,y)=E_{xx}(x,y)\ , \quad K(x,y)=-E_{xy}(x,y)\ , \quad 
U(x,y)=-\sfrac13\bigl( E_{xx}(x,y)+ E_{yy}(x,y)\bigr)\ ,
\ee
which is derived from the generating function $E(x,y)$ obeying the master equation (\ref{ME}).

Note that in practice, in order to verify that the system (\ref{EqM}) admits an extra cubic 
integral of motion (\ref{I2cub}), it suffices to substitute $J(x,y)$, $K(x,y)$ and $U(x,y)$ 
into (\ref{KJU}) and to verify that the equations are satisfied identically. 
As an exercise, one can check that the models exposed below in (\ref{samples}) 
indeed obey the structure equations (\ref{KJU}). 
Of course, finding the explicit form of the potential $U$ giving rise to a cubic integral
of motion requires solving the master equation~(\ref{ME}). 
This will allow us to construct new two-dimensional integrable models.

A similar situation holds for the case of a polynomial quartic in the velocities,
\be\label{QI}
I_2\={\dot x}^4+P(x,y)\,{\dot x}^2+Q(x,y)\,\dot x\,\dot y+R(x,y)\ .
\ee
For this to be an integral of motion of the system (\ref{EqM}), the partial differential equations
\bea\label{PQRU}
&&
Q_x+P_y\=0\ , \qquad \quad P_x-Q_y+4 U_x\=0\ ,
\nonumber\\[2pt]
&&
R_y+U_x Q\=0\ ,\qquad R_x +U_y Q+2 U_x P+2 Q_y U\=0\
\eea
must be obeyed~\cite{Y}.\footnote{ 
Note that, in deriving (\ref{PQRU}), the zero energy condition was used again 
to express ${\dot y}^2$ via ${\dot x}^2$ and $U$.}
Repeating the same arguments as above, one can verify that (\ref{PQRU}) is equivalent 
to the chain of relations~\cite{Y}
\bea\label{sf}
&&
P(x,y)\=F_{xx}(x,y)\ , \qquad
Q(x,y)\=-F_{xy}(x,y)\ , \qquad U\=-\sfrac 14 \bigl( F_{xx}+ F_{yy}\bigr)\ ,
\nonumber\\[2pt]
&&
R(x,y)\=-\int_{y_0}^y\!\diff\tilde{y}\ \bigl(Q\,U_x\bigr)(x,\tilde y)  -
\int_{x_0}^x\!\diff\tilde{x}\ \bigl( Q\,U_y +2 P\,U_x+2 U\,Q_y\bigr)(\tilde x,y_0)\ ,
\eea
which all are derived from the single generating function $F(x,y)$ obeying the master equation
\be\label{me}
(F_{xxxx}F_{xy}-F_{yyyy}F_{yx})+3(F_{xxx}F_{xxy}-F_{yyy}F_{yyx})+2(F_{xx}F_{xxxy}-F_{yy}F_{yyyx})\=0\ .
\ee
Again, it should be stressed that, given the polynomial (\ref{QI}), 
the partial differential equations~(\ref{PQRU}) allow one to directly verify that (\ref{QI}) 
is an integral of motion of the system (\ref{EqM}), 
irrespective of any knowledge of the generating function $F(x,y)$.
As an example, one may consider the system given below in (\ref{ExQ}) and check that (\ref{PQRU}) holds. 
Yet, for defining new models possessing a quartic integral of motion we have to find 
a generating function~$F$ solving~(\ref{me}).

It is unknown how to solve the master equations (\ref{ME}) or~(\ref{me})
in full generality. So far, particular solutions have been constructed
in a form where the variables are separated~\cite{Sel,Sel1,HS,Y,Y1},
\be\label{ys}
E(x,y) \quad\textrm{or}\quad F(x,y)\=H_1(x)+H_2(y)+\Psi(x)\,\Phi(y)\ .
\ee
This turns out to encompass almost all known models with a cubic or quartic
second integral of motion.
Our goal in this work is to construct new solutions to (\ref{ME}) and~(\ref{me})
which are not of the type~(\ref{ys}) and thus generate new models with a cubic or a quartic integral of motion.

In Section~2 we analyze the symmetries of the master equations and uncover the dihedral groups 
$D_6$ and $D_8$ for the cubic and quartic cases, respectively. 
The dihedral symmetry suggests the use of special invariant variables, which allow one 
to reduce the master equations (\ref{ME}) and~(\ref{me}) to nonlinear ordinary differential equations.  
It will be shown that the reduced master equation for the quartic case is just the derivative 
of the corresponding equation in the cubic case.
In Section~3 the reduced master equation for the cubic case is solved in full generality, 
which also provides a particular solution for the quartic case. 
The corresponding integrable models are discussed, and the most simple examples are displayed explicitly. 
The concluding Section~4 contains the summary and an outlook.  

\vskip 0.5cm

\noindent
{\bf 2.  Symmetries of the master equations and invariant variables}

\vskip 0.3cm
\noindent
Before making some ansatz, it is advisable
to investigate the symmetries of the master equations (\ref{ME}) and~(\ref{me}).
Those may reveal adapted coordinate systems in which the analysis simplifies.
Obviously, we can independently rescale the functions or both coordinates $x$ and~$y$,
but the latter two simultaneously.
Hence, with any solution $E(x,y)$ or $F(x,y)$ we immediately have a two-parameter family
\be\label{scaling}
E^{\lambda,\mu}(x,y)\=\lambda\,E(\sfrac{x}{\mu},\sfrac{y}{\mu}) \und
F^{\lambda,\mu}(x,y)\=\lambda\,F(\sfrac{x}{\mu},\sfrac{y}{\mu})
\qquad\textrm{with}\qquad \lambda,\mu\in\R\ .
\ee
But there is more:
Let us rewrite the master equations in terms of complex coordinates
\be
z=x+\ic y \and \bz=x-\ic y \qquad\Rightarrow\qquad
\pa_x=\pa_z+\pa_{\bz} \and \pa_y=\ic(\pa_z-\pa_{\bz})\ ,
\ee
which produces
\bea
&& \pa_z\bigl(E_{zz}E_{z\bz}\bigr)\ +\ \pa_{\bz}\bigl(E_{\bz\bz}E_{\bz z}\bigr)\= 0
\und \\[6pt]
&& \pa_z\bigl(2F_{zz}F_{zz\bz}+F_{zzz}F_{z\bz}\bigr)\ -\
\pa_{\bz}\bigl(2F_{\bz\bz}F_{\bz\bz z}+F_{\bz\bz\bz}F_{\bz z}\bigr)\= 0\ .
\eea
It is apparent that these equations are invariant under
\be
z\ \mapsto\ \eu^{\ic\alpha}z \qquad\textrm{for}\qquad
\alpha\=\sfrac{\pi}{3} \and \alpha\=\sfrac{\pi}{4}\ ,
\ee
respectively. Together with their invariance under the conjugation $z\mapsto\bz$,
the symmetry transformations generate the dihedral group $D_6$ respectively $D_8$
(the symmetry group of the 6-gon respectively 8-gon), with 12 respectively 16 elements.
Below we treat the two cases in turn.

For the cubic situation in the $(x,y)$ coordinates, the $D_6$ group is generated by
\be\label{D6}
(x,y)\ \mapsto\ \sfrac12(x{-}\sqrt{3}y,\sqrt{3}x{+}y) \und (x,y)\ \mapsto\ (-x,y)\ .
\ee
The sign of $y$ may also be flipped independently.
As a consequence, every solution~$E$ produces 12 other solutions by discrete transformations,
some of which may coincide if~$E$ is invariant under part of the $D_6$ group.
{}From these observations, it is clear that
\be\label{trivial2}
E(x,y) = E_1(y) \qquad\textrm{or}\qquad
E(x,y) = E_2(\sqrt{3}x{+}y) \qquad\textrm{or}\qquad
E(x,y) = E_3(\sqrt{3}x{-}y)
\ee
yield trivial solutions.
Let us search for solutions $E$ invariant (possibly up to sign)
under part of the $D_6$ symmetry.
To be more precise, we identify $D_6$ invariant combinations
of $x$ and~$y$ as improved coordinates and suggest to use them in searching for solutions 
of the master equation (\ref{ME}).  These are obtained by multiplying
the original coordinates with all their $D_6$ images, which yields
(up to overall coefficients)
\be
x^2(x{-}\sqrt{3}y)^2(x{+}\sqrt{3}y)^2 \und y^2(\sqrt{3}x{+}y)(\sqrt{3}x{-}y)^2\ .
\ee
Weakening the demand to $D_6$ invariance only up to a sign, it suffices to take
\be\label{VarU}
v\=x^3-3x\,y^2 \und
u\=3x^2y-y^3 \ .
\ee

{}From (\ref{trivial2}) we see that $E_1(y)$ solves~(\ref{ME}) while $E_1(x)$
does not. Therefore, we look for solutions which depend only on~$u$,
\be\label{an2}
E(x,y)\=f(u)\ .
\ee
Substitution in (\ref{ME}) yields the nonlinear ordinary differential equation
\be\label{me2}
f' f^{(3)}+3u f'' f^{(3)}+ 4 f'' f'' \=0 \ ,
\ee
where the prime denotes the derivative with respect to $u$.
Since $f$ appears only via its derivatives, we pass to
\be
p(u)\=f'(u)
\ee
and bring (\ref{me2}) to the form
\be\label{int2}
p\,p''+3u\,p' p''+4\,p'p'\=0\ .
\ee
We note that the potential
\be\label{Pot}
U(x,y)\= -3(x^2{+}y^2)^2\,p'(u)
\ee
and the building blocks
\be\label{BuilBl}
J(x,y)\= -6 y p(u)+36 x^2 y^2 p'(u) \und
K(x,y)\= 6x p(u)-18xy (x^2{-}y^2)\,p'(u)
\ee
are constructed directly from the solution~$p(u)$. 
Thus we have demonstrated that solving the nonlinear ordinary differential equation (\ref{int2})
will produce two-dimensional integrable models admitting a cubic integral of motion.

Before solving (\ref{int2}) in full generality, let us repeat the exercise for the quartic case.
In the $(x,y)$ coordinates, the basic $D_8$ actions read
\be\label{D8}
(x,y)\ \mapsto\ \sfrac1{\sqrt{2}}(x{-}y,x{+}y) \und (x,y)\ \mapsto\ (-x,y)\ .
\ee
They generate, in particular, the interchange of $x$ and~$y$ and
independent sign flips of $x$ or~$y$.
Hence, every solution~$F$ produces up to 15 other solutions by discrete transformations, 
some of which may coincide.
Note that
\be\label{trivial}
F(x,y) \= F_1(x)+F_3(y) \qquad\textrm{or}\qquad
F(x,y) \= F_2(x{-}y)+F_4(x{+}y)
\ee
yields trivial solutions.\footnote{
At the end of the paper we comment more on such cases.}

Let us look for solutions $F$ invariant (possibly up to sign) under part of the $D_8$ symmetry.
To this end, we identify more symmetric combinations of the variables by multiplying suitable
$D_8$ images of $x$ or~$y$.
The simplest possibilities are the so-called parabolic coordinates
\be\label{VarS}
s\= x\,y \und t\= \sfrac12(x^2{-}y^2)\ ,
\ee
which are invariant under
$(x,y)\mapsto(-x,-y)$, so that only a $D_4$ subgroup remains effective.
Indeed, the two generators in~(\ref{D8}) act as
\be \label{D8st}
(s,t)\ \mapsto\ (t,-s) \und (s,t)\ \mapsto\ (-s,t)\ .
\ee
Consider solutions which depend only on one of the parabolic variables,
thus being manifestly invariant under a quarter of the original $D_8$ symmetry.
{}From the $x\leftrightarrow y$ symmetry of~(\ref{me}) it does not matter
which variable we choose, so we consider the ansatz
\be\label{an}
F(x,y)\=f(s)\ ,
\ee
which is inert under $(x,y)\mapsto(y,x)$ as well.
Substitution in (\ref{me}) yields the nonlinear ordinary differential equation
\be\label{me1}
f'f^{(4)}+3s f''f^{(4)}+12f''f^{(3)}+3s f^{(3)} f^{(3)}\=0 \ ,
\ee
where the prime now denotes the derivative with respect to $s$.
It can immediately be integrated once to
\be
f' f^{(3)}+3s f'' f^{(3)}+ 4 f'' f'' \= C
\ee
with an integration constant~$C$.
Since $f$ appears only via its derivatives, we pass to
\be
p(s)\=f'(s)
\ee
and bring (\ref{me1}) to the form
\be\label{int}
p\,p''+3s\,p' p''+4\,p'p'\=C\ .
\ee
We note that the building blocks
\be\label{BBQ}
U(x,y)\= -\sfrac14(x^2{+}y^2)\,p'(s)\ ,\qquad
P(x,y)\= y^2 p'(s)\ ,\qquad
Q(x,y)\= -(1+s\pa_s)p(s)
\ee
and $R(x,y)$ are constructed directly from the solution~$p(s)$. 
Solving (\ref{int}) will thus provide integrable models with a quartic integral of motion.

It will not have escaped the reader's attention that, for vanishing integration
constant~$C$, the reduced master equation~(\ref{int}) coincides with the reduced
master equation~(\ref{int2}) for the cubic case!
Therefore, the cubic task is contained in the quartic one, and we may again
discuss them together from now on, writing $s$ as the variable in both cases.

The reduced master equation~(\ref{int}) is still invariant under the following two
continuous transformations,
\be\label{pscale}
\bigl(\tilde{p}(s)\,,\,\tilde{C}\bigr) \= \bigl(\lambda\,p(s/\lambda)\,,\,C\bigr) \und
\bigl(\tilde{p}(s)\,,\,\tilde{C}\bigr) \= \bigl(\lambda\,p(s)\,,\,\lambda^2 C\bigr)\ ,
\ee
which yield new solutions $\tilde{p}$ from an old solution~$p$.
The first case amounts to a simultaneous rescaling of $p$ and $s$, keeping $p'$
and $C$ inert, while the second case scales only~$p$ but also changes~$C$,
relating solutions for different (nonzero) values of the integration constant.
Therefore, it suffices to consider $C=0,+1,-1$ only.
Unfortunately, we failed to solve (\ref{int}) for nonzero~$C$.\footnote{
One particular $C{\neq}0$ solution gives a Calogero plus harmonic potential.}
So in what follows we restrict ourselves to the case $C=0$,
which is no loss in the cubic case and keeps both
transformations in~(\ref{pscale}) viable.

\newpage

\noindent
{\bf 3.  Reduced master equation and new integrable models }

\vskip 0.3cm
\noindent
The key to solving
\be\label{int0}
p\,p''+3s\,p' p''+4\,p'p'\=0
\ee
is the substitution
\be\label{pp}
p'(s)\=s^{-1} g\bigl(p(s)\bigr)\ ,
\ee
where $g(p)$ is a function to be determined. This reduces (\ref{int0}) to the form
\be\label{supp}
0 \= \bigl(p+3g(p)\bigr)\frac{\diff g}{\diff p}(p)+g(p)-p
\=\frac12\frac{\diff}{\diff p}\Bigl[\bigl(3g(p)-p\bigr)\bigl(g(p)+p\bigr)\Bigr]\ ,
\ee
from which we obtain, with an integration constant written as~$\frac43\beta^2$ for later convenience,
\be\label{gsol}
\bigl(3g(p)-p\bigr)\bigl(g(p)+p\bigr)\=\sfrac43\beta^2
\qquad\Rightarrow\qquad
g(p) \= -\sfrac13\bigl( p + 2\epsilon\sqrt{p^2+\beta^2}\bigr)\ ,
\ee
where $\epsilon=\pm1$ distinguishes the two roots of the quadratic equation.
Inserting~(\ref{gsol}) into~(\ref{pp}), we get an ordinary differential equation
which is readily solved by
\be
s\= \bigl( p + 2\epsilon\sqrt{p^2+\beta^2}\bigr)\bigl(p+\sqrt{p^2+\beta^2}\bigr)^{-2\epsilon}\ ,
\ee
where an integration constant was absorbed into rescaling~$s$.
The scaling freedom~(\ref{pscale}) also allows us to reduce $\beta^2$ to $\pm1$ unless it is zero.

To invert for $p(s)$ at $\beta^2{=}\pm1$, we bring this relation to the form
\be
4\epsilon\,s\,p^3 -3\,p^2 \pm 6\epsilon\,s\,p + s^2\mp 4 \= 0\ ,
\ee
which shows that the two cases $\epsilon=\pm1$ differ simply by a sign flip of $s$.
A solution to this cubic equation reads~\footnote{
For the upper sign choice, it is the unique real solution.
For the lower sign choice, we can always define a real branch of the cubic roots.
In the interval $s\in[-1,1]$, two further real solutions exist.}
\be
p(s) = \frac{\epsilon}{4s}\Bigl\{ 1 + (1\mp8s^2)
\bigl[1\pm20s^2-8s^4+8\epsilon s(s^2\pm1)^{\frac32}\bigr]^{-\frac13} +
\bigl[1\pm20s^2-8s^4+8\epsilon s(s^2\pm1)^{\frac32}\bigr]^{+\frac13} \Bigr\}\ .
\ee
The full two-parameter family is obtained by reinstating the scaling variables,
\bea
&& p(s) \= \epsilon\frac{\lambda\mu}{4s}\Bigl\{ 1 + (\mu^2\mp8s^2)
\bigl[\mu^6\pm20\mu^4s^2-8\mu^2s^4+8\epsilon\mu^2s\,(s^2\pm\mu^2)^{\frac32}\bigr]^{-\frac13}
\Bigr.\nonumber \\ &&\qquad\qquad\qquad\qquad\qquad\!\! \Bigl. +\ \mu^{-2}
\bigl[\mu^6\pm20\mu^4s^2-8\mu^2s^4+8\epsilon\mu^2s\,(s^2\pm\mu^2)^{\frac32}\bigr]^{+\frac13}
\Bigr\}\ . \label{psol}
\eea
In order to construct the corresponding integrable models, its suffices to substitute this function
into the building blocks (\ref{Pot}), (\ref{BuilBl}) and (\ref{BBQ}) for the cubic and quartic cases, 
respectively, and to express the variables $u$ and $s$ in terms of $x$ and $y$ via (\ref{VarU}) 
and (\ref{VarS}). The advantage of our models is that they are given in terms of elementary functions 
(cf.~\cite{Y}).

Because the resulting models are rather bulky, its is instructive to dwell on the special case 
of $\beta=0$. The two possibilities $\epsilon=\pm1$ yield
\bea
&& g(p)\=-p \qquad\Rightarrow\qquad p(s)\=\lambda s^{-1} \ ,\\[6pt]
&& g(p)\=\sfrac13 p \qquad\;\Rightarrow\qquad p(s)\=\lambda s^{\frac13}\ ,
\eea
We remark that for small and for large values of~$s$ the function $p$ in~(\ref{psol})
asymptotes to the first and the second $\beta{=}0$ solutions, respectively.
Let us expose the potential and the higher invariant for $\beta=0$.
In the cubic case, one finds
\bea\label{samples}
&&U= \frac{3 \lambda {(x^2{+}y^2)}^2}{{(3x^2y{-}y^3)}^2} \und
I_2={\dot x}^3-\frac{6\lambda(3x^2{+}y^2)}{{(3x^2{-}y^2)}^2}\dot x-
\frac{12\lambda xy}{{(3x^2{-}y^2)}^2} \dot y \ ,\\[8pt]
&&U=-\frac{\lambda {(x^2{+}y^2)}^2}{{(3x^2y{-}y^3)}^{\frac23}} \und
I_2={\dot x}^3+\frac{6\lambda y^2(5x^2{-}y^2)}{{(3x^2y{-}y^3)}^{\frac23}}\dot x
-\frac{12\lambda xy(2x^2{-}y^2)}{{(3x^2y{-}y^3)}^{\frac23}} \dot y
\nonumber
\eea
for $\epsilon=+1$ and $\epsilon=-1$, respectively. 
It is straightforward to verify that the structure functions $U$, $J$ and $K$ derived from (\ref{samples}) 
obey the equations (\ref{KJU}), which provides a consistency check for our considerations.
The quartic case produces
\bea\label{ExQ}
U= -\sfrac{\lambda}{12} (x^2{+}y^2)(xy)^{-\frac23} \and
I_2=\dot{x}^4+\sfrac{\lambda}{3}(xy)^{\frac13}\sfrac{y}{x}\dot{x}^2
-\sfrac{4\lambda}{3}(xy)^{\frac13}\dot{x}\dot{y}
-\sfrac{\lambda^2}{36}(xy)^{\frac23}\bigl(8{-}\sfrac{y^2}{x^2}\bigr)
\eea
for $\epsilon=-1$, while $\epsilon=+1$ yields merely two decoupled conformal particles in one dimension. 
Again, one readily verifies that the functions $U$, $P$, $Q$ and $R$ derived from (\ref{ExQ})
solve (\ref{PQRU}).
To the best of our knowledge, the models (\ref{samples}) and (\ref{ExQ}) are new.
Acting with $D_6$ respective $D_8$ transformations on these solutions yields little new besides
the obvious possibility of replacing $s$ with~$t$ in all quartic-integral solutions.

We conclude with a remark concerning the second form  of a trivial ansatz in~(\ref{trivial}).
Clearly, $F=F_2+F_4$ satisfies the two-dimensional wave equation,
\be\label{we}
F_{xx}-F_{yy}\=0\ ,
\ee
which leads to the simplifications
\be\label{mis}
P(x,y)\=-2 U(x,y) \und  R(x,y)\=-\sfrac 14 Q(x,y)^2
\ee
and finally to
\be\label{mis1}
I_2\=-\bigl(\dot x\,\dot y-\sfrac12 Q(x,y)\bigr)^2\ .
\ee
Thus, the quartic integral is the square of a quadratic one and, hence, it is reducible.
As an illustration, let us consider the following polynomial of fifth degree,
\bea
&& F(x,y)\=\bigl\{\sfrac1{60}(x{+}y)^5+\sfrac1{24}(x{+}y)^4\bigr\}
-\bigl\{\sfrac1{60}(x{-}y)^5-\sfrac1{24}(x{-}y)^4\bigr\} \nonumber \\[8pt]
&& \qquad\quad\
\=\sfrac1{60}\bigl( 5x^4+10x^4y+30x^2y^2+20x^2y^3+5y^4+2y^5\bigr)\ .
\eea
In this case, the potential and the quartic integral of motion read
\be
U(x,y)\=-\sfrac12 (x^2{+}y^2)-x^2 y-\sfrac13 y^3 \und
I_2\=-\bigl(\dot x\,\dot y+x\,y+\sfrac13 x^3+x\,y^2)^2\ ,
\ee
which reproduces the model studied in~\cite{aiz}.\footnote{
Our convention for the potential differs from theirs by a flip of sign.}

\vskip 0.5cm

\noindent
{\bf 4.  Summary and outlook}

\vskip 0.3cm
\noindent
In this work we uncovered a dihedral $D_6$ or $D_8$ symmetry of the master equation
underlying two-dimensional integrable models featuring a second integral of motion
which is cubic or quartic in the velocities. We introduced a symmetry-adapted ansatz
employing invariant variables, which reduced the master equation to an ordinary 
differential equation. The latter was then solved by conventional methods. 
Interestingly, the reduced master equation for the quartic case is just the derivative 
of the corresponding equation in the cubic case.
Our second result is that solutions of the two-dimensional wave equation
generate another family of integrable models possessing a quartic invariant which,
however, degenerates to the square of a quadratic one and, thus, is reducible.
Finally, we conjecture that, quite generally, the master equation governing
two-dimensional models with a constant of motion of $n$th order in the velocities
enjoys a $D_{2n}$ invariance. We hope to test and apply this idea to the quintic and
sextic cases in the future.

\vspace{0.5cm}

\noindent{\bf Acknowledgements}

\vspace{0.3cm}
\noindent
A.G. is grateful to the Institut f\"ur Theoretische Physik at Leibniz Universit\"at Hannover
for the hospitality extended to him during the course of this work.
The research was supported by the DFG grant LE 838/12-1 and RFBR grant 13-02-91330.


\end{document}